\documentclass[onecolumn,showpacs,preprintnumbers,amsmath,amssymb,aip]{revtex4}
\usepackage{graphicx}
\usepackage{dcolumn}
\usepackage{bm}
\begin{document}

\title[ Anomaly in  Andreev reflection  at large FM/high-T$_{C}$ superconductor  ]{  Anomaly in  Andreev reflection at large ferromagnet/high-T$_{C}$ superconductor  area junctions with rough interface }


\author{ N. Ghosh$^{1,*}$ H. Schmidt$^{2}$ and M. Grundmann$^{1}$  \\
 {\small $^{1}$Institut f{\"u}r Experimentelle Physik II, Fakult{\"a}t f{\"u}r Physik und Geowissenschaften,
 Universit{\"a}t Leipzig, Germany}\\
{ \small $^{2}$Forschungszentrum Dresden-Rossendorf, Institut f{\"u}r lonenstrahlphysik und Materialforschung, Germany}}

\date{\today}

\begin{abstract}
\normalsize \baselineskip=18pt

Sub-gap conductance at a large area junction  with a rough interface of a  ferromagnet and  a high-T$_{C}$  superconductor  is superimposed by multiple peaks which is not expected from an ideal point contact Andreev reflection process. We demonstrate this phenomenon by measuring resistance as a function of bias voltage of a Co/Y$_{1}$Ba$_{2}$Cu$_{3}$O$_{7-\delta}$ junction with contact area  50 x 70 $\mu$ $m^{2}$ at various temperatures. In order to analyze  such Andreev reflection data, the interface is assumed to have  random potentials which can create local electric fields. The Blonder-Tinkham-Klapwijk theory is modified with the inclusion of a broadening parameter due to finite life time effects of quasi particles. An additional voltage drop due to local electric fields at the rough interface has been included in terms of an extra energy shift which may be related to the asymmetry of normalized resistance data. Spin polarization has been introduced for the ferromagnet. The presented model  explains the multi-peak nature and asymmetry of Andreev reflection data experimentally observed at large area junctions. Extension of the model also interprets the experimentally observed anomalous enhancement of resistance peaks in the sub-gap  region which may result from crossing  the critical current limit across the junction.


\end{abstract}

\maketitle

\newpage
\section{Introduction}
Andreev reflection (AR) is a process that converts quasi-particle currents in a normal metal (N) to supercurrents in a superconductor (SC) at a N/SC-  interface. Only a single electron  in N with energy (E) smaller than the superconducting gap ($\Delta$) can get into the SC  by pairing up with an electron of opposite momentum and spin. This is accomplished by the reflection of a hole back
into the metal, a process first described by Andreev.\cite{af} AR  results in an enhancement of the N/SC junction conductance beyond its normal state value in the sub-gap region. The theoretical explanation of AR was given by  Blonder,  Tinkham and  Klapwijk  in 1982.\cite{BTK} When N is replaced by a ferromagnet (FM), AR  at the FM/SC interface is suppressed due to the spin imbalance leading to a decrease of the sub-gap conductance. A measurement of this suppression  yields the magnitude of spin polarization ($\it{P}$) in the FM.\cite{soulen,upa} In order to incorporate spin polarization in the modelling, Strijkers  $\it{et\ al}$.  introduced a method  of subdividing the total current across the interface into  a polarized and  an unpolarized part.\cite{Strijker} The determination of  $\it{P}$ by modelling AR spectroscopic data  is strongly dependent on the interface quality of the FM/SC junction. The Blonder-Tinkham-Klapwijk (BTK)  theory describes  the interface as a  delta  function potential.\cite{BTK} This is a simple description and it can not describe AR at rough large area junctions. Hence, modification is necessary in order to explain realistic experimental AR data. \\
Depending on the interface quality, the junction resistance also varies. Classical  junctions of macroscopic size are governed by Maxwell resistance ($\it{R_{M}}$) for a small contact area of radius $\it{d}$ $\ll$ $\it\lambda$ with $\it\lambda$  the mean free path and  $\it\rho$  the specific resistivity of a metal, $\it{R_M = \rho/2d}$. Ballistic point contacts  are defined in a semi-classical description for $\it{d}$ $\gg$ $\it\lambda$  and are governed by Sharvin resistance $\it{R_{S} = \frac {4\lambda }{ 3 \pi d^{2}}}$. Wexler's formula relates Maxwell and Sharvin resistance by $\it{ R_{contact} = R_{S} + G(\lambda/d) R_{M}}$, where $\it{G(\lambda/d)}$ is a slowly varying function. Essentially,  the BTK theory is applicable to  ballistic point contacts and  s-wave superconductors.
A tunneling theory for normal metal-insulator-$\it{d}$-wave
superconductors has been reported by Tanaka and Kashiwaya.\cite{tana_kashi} However, it has been shown by Barash $\it{et\
al.}$ \cite{Barash} that the character of the change in the order
parameter at the junction of a $\it{d}$-wave superconductor and
normal metal is not much  different from a junction with
a $\it{s}$-wave superconductor when the orientation of the normal metal to
the $\it{d}$-wave superconductor is along the principal crystallographic axes. It implies that AR at
 a junction of N and $\it{d}$-wave SC with junction normal along a principal crystallographic axes of $\it{d}$-wave SC  is similar to AR at the isotropic  N/s-wave SC junction.  For example, a recent AR study reveals that when  a thin  Y$_{1}Ba_{2}Cu_{3}O_{7-\delta}$(YBCO) film, a high-T$_C$ superconductor, is oriented
with the c-axis normal to the interface, a zero bias conductance peak is observed.\cite{luo} These results evidence
that the problem of anisotropy in the order parameter can be solved by  studying  FM/high-T$_{C}$  SC junctions along a principal crystallographic axes. However, the interface roughness is difficult to be avoided. For example, it is reported that YBCO films grow with
large-angle grain boundaries and obstruct the development of reproducible Josephson junctions  of different types.\cite{alex} Droplet formation on YBCO thin films grown by pulsed laser deposition is a known feature \cite{mozhaev} and an unavoidable source of noise which occurs during conductance measurements across FM/YBCO junctions. The interface can never be  smooth and homogeneous,if a normal or ferromagnetic material is deposited on an amorphous superconductor like tungsten carbide \cite{ralph}. The roughness of surface has already been taken into account to do quantitative fitting of AR data in high-T$_{C}$  SC.\cite{Fogel}  Hence, it is necessary  for us to consider  the influence of interface scattering and roughness  when modelling real AR thereby  modifying initial BTK formula.\\
Additionally, when the current across the FM/high-T$_{C}$  SC  junction reaches  locally the critical current ($\it{I_{C}}$) for the superconductor at any bias (eV), the superconductor becomes a normal metal. For a fixed bias the critical current may be locally reached at a rough FM/SC junction yielding an increase of resistance where the superconductivity is lost, thus causing additional noise in the differential resistance measurement.\cite{goutam,hind} As a result, AR data are superimposed by heavy noise, when AR measurements are carried out on rough FM/SC junctions. In general, noise is a time dependent phenomenon. However, in the present case, noise refers to the multi-peak nature of differential resistance data as a function of bias voltage. We extend the original BTK model in order to explain  the anomalous increase of resistance caused by critical current effects also.    \\

In this paper, we  report on the  AR  experiments and modelling  of  AR data across FM/high-T$_{C}$  SC  junctions grown along the (001) crystallographic direction. The motivation  is to identify the problems in Andreev reflection usually occurs in large area FM/high-T$_{C}$ junction and to try to explain them in the light of modified BTK modelling. We have dealt  mainly with the problems faced during the  AR experiments   with FM/high-T$_{C}$  SC samples. Efforts have been made to theoretically  model those  AR data  with help of the relatively new idea of multi-point contacts. Since surface roughness is inevitable  for high-T$_{C}$ SC like YBCO, we believe  large N-metal/YBCO   junction would be the right candidate for invoking this new concept. In addition, we chose FM-metal/YBCO junction  in order to show that  we can  extract  information about spin polarization of  FM  by  modeling AR with multi-point contacts.
We demonstrate  the occurrence of anomalous  multi-peaks  in  AR data by carrying out resistance measurements as a function of  bias voltage for Co/YBCO samples with 50x70  $\mu$m$^{2}$ contact area. We will show how the interface roughness influences the shift of   the resistance dip away from zero bias. Our model is capable of determining the $\it{P}$ of Co and the $\Delta$ of YBCO. Furthermore, we will discuss the additional  enhancement of resistance due to the influence of critical current (Sect. 4.1).\\
\section{Theory}
\subsection{Ideal Andreev reflection}
The BTK model employs a generalized semiconductor scheme to match the wavefunctions at the N/SC interface with some specific boundary conditions. The system we consider is described by  Bogoliubov-de-Gennes   matrix equations  for electrons and  holes which are coupled.  Because the point contact is assumed to be along a crystallographic axis, the model may be restricted to one dimension in the present case of small contact area. The repulsive potential   H $\delta$ (x)  models the elastic scattering  that usually occurs in the orifice at the N/SC interface.  The diameter and the width of the interface should be less than the mean free path $\it\lambda$  of N and the coherence length $\xi$ of  SC. The interface  barrier strength is varied  through a parameter $\it{Z}$ = H/($\hbar$ v$_{F}$), where v$_{F}$ is the Fermi velocity.
According to BTK theory, the current (I$_{NS}$) across a N/SC junction can be  described as,

\begin{equation}
I_{NS}= 2e N v_{f} \mathcal{A} \int_{\infty}^{\infty}[f(eV)][1+A(E)-B(E)]dE,
\end{equation}
where
\begin{eqnarray*}
A(E)= \frac{\Delta^{2}}{E^{2}+ ( \Delta^{2}-E^{2})(1 + 2Z^{2})},& B(E) =1- A(E),\\
f(eV) = f_{0}(E-eV)-f_{0}(E)&
\end{eqnarray*}
and,  e=electronic charge, N = BCS density of states, v$_{F}$ = Fermi velocity, $\mathcal{A}$= junction
cross sectional area, f$_{0}$ = Fermi distribution function, V =
applied voltage, A(E)= probability of Andreev reflection and B(E)=
probability of normal reflection. Temperature effects are included via v$_{F}$ and f$_{0}$.\\
It is well known that Cooper pairs of the superconductor layer usually get diffused into the metal and a weakly superconducting layer is created in the metallic side of the junction. This is called proximity effect.
 On the microscopic level, the proximity effect is completely described by the Andreev reflection process.
This superconducting proximity layer has a lower transition temperature and a lower $\it\Delta$ than  the bulk superconductor. The AR  occurs
at the metal/proximity-layer interface and it is limited to bias voltages smaller than the superconducting gap value ($\it\Delta_{1}$) of the proximity layer. However, quasiparticles can only enter the SC for voltages higher than the bulk gap($\it\Delta_{2}$).
Hence, there are two gap parameters to be incorporated into the fitting procedure, one for Andreev reflection and proximity effect  $\it\Delta_{1}$,  the other for quasiparticle transport $\it\Delta_{2}$.\cite{Strijker} We have observed that the modelled $\it\Delta_{1}$  is always smaller than $\it\Delta_{2}$ .\\

\subsection{Effect of spin polarization}

The BTK theory is valid  for AR at a N/SC junction. This model does not give information about spin polarization when the metal is ferromagnetic. In order to account for the influence of $\it{P}$ on AR, we have to consider that when a spin-up polarized electron is incident at the junction, it needs another spin-down polarized electron to get into the SC. Because a  Cooper pair in the superconductor is composed of  a  spin-up and a spin-down  electron, the removal of the spin-down
electron leaves a spin-up hole which is Andreev
reflected back into the metal. Since,
the spin-up hole is considered as the absence of a spin-down
electron and so by convention it should be in the spin-down
density of states (DOS).\cite{soulen} The magnitude of AR  changes in dependence on the availability of spin-down electronic states.
 For example, there will be no spin-down electronic states for a 100$\%$ spin polarized material and AR is not observed. The effect of $\it{P}$ can be taken into account by separating the total current I$_{NS}$ across the junction into  polarized I$_{P}$  and unpolarized I$_{U}$ parts. \cite{Strijker} We have substituted the expression of  I$_{NS}$ from Eq.~1 for I$_{U}$ and I$_{P}$,  differentiated  I$_{NS}$ with respect to voltage  as  given below  and then  numerically integrated on energy scale.

\begin{equation}
dI_{NS}/dV   =  (1-P) dI_{U} /dV +  P  dI_{P} /dV
\end{equation}

where \\

\begin{eqnarray*}
dI_{U}/dV &=& \frac{1}{R_{N}}\int_{\infty}^{\infty}\frac{df(eV)}{d(eV)}[1+A_{U}(E)-B_{U}(E)]dE\\
dI_{P}/dV &=& \frac{1}{R_{N}}\int_{\infty}^{\infty}\frac{df(eV)}{d(eV)}[1+A_{P}(E)-B_{P}(E)]dE\\
R_{N} &=& \frac{1 + Z^{2}}{2e^{2}Nv_{f}\mathcal{A}}
\end{eqnarray*}

\subsection{ Influence of interface scattering}
Because the modelled  conductance peaks are  sharp and do not match  the broad  experimental conductance, sometimes it is not sufficient to model AR data with the parameters, $\it{Z}$,  $\it\Delta_{1}$, $\it\Delta_{2}$, and $\it{P}$. Here, we take into consideration   broadening  of   peaks   by   including  a  complex  part  in the energy  as  $E^{'} = E + i\it\Gamma$   according  to P. Szabo $\it{et\ al.}$  where  large scattering  at the interface and finite life time effects are  taken  into account by $\it\Gamma$.\cite{Szabo} This broadening parameter has been  proposed \cite{Dynes} and employed by several authors
before. \cite{Srikanth} The BCS density of states in Eq.~1  $N_{S}(E)$( $|E|/(E^{2}-\Delta^{2})^{1/2})$ is modified  to  $N^{'}_{S}(E)$
($ \Re[|E-i\Gamma|/((E-i\Gamma)^{2}-\Delta^{2})^{1/2}$]) when the lifetime broadening parameter $\it\Gamma$ is included. It should be mentioned that the BTK model is restricted to one dimension. On a scale shorter than the coherence length $\xi$, the energy gap ($\it\Delta$) rises to its asymptotic value and  the evanescent waves also decay.\cite{BTK} In the case of high-T$_{c}$ superconductors most likely
this is violated due to extremely short coherence lengths. Moreover, the existence of a thin native non-superconducting layer  on YBCO  due to surface degradation has been probed by tunneling measurements.\cite{Iguchi} This gives rise to inelastic scattering to the incoming electrons. Usually, such normal state scattering destroys the quantum phase coherence
of the carriers crossing the junction and the modification of BTK model incorporating  finite life time broadening is quite relevant. Hence, the inclusion of an imaginary term in the energy is an effective technique to describe life time effects in tunneling and it is reasonable to do the same in the BTK framework also. The modelling  with  five parameters, $\it{Z}$, $\it\Delta$($\it\Delta_{1}$, $\it\Delta_{2}$), $\it{P}$ and $\it\Gamma$, is  quite useful for  sub-gap conductance data  probed on  FM/SC junctions. The interface scattering can be efficiently taken into account by  the $\it\Gamma$ parameter.


\section{Experiments}
YBCO films have been deposited  by pulsed laser deposition on sapphire substrates in  oxygenated atmoshphere.\cite{lorenz} The Co/YBCO heterostructure has been prepared by thermal evaporation of Co on YBCO under a vacuum level of 3 x 10$^{-3}$ mbar.
Actually,  the  Co deposition was carried out on YBCO surface in ex-situ condition. Since, YBCO surface was  exposed in air before, the oxygen atoms have  been adsorbed at apical cites  as described by M. Naito.\cite{naito}  If the  Co took oxygen from the  YBCO surface, it would have  taken only the adsorbed oxygen. Hence, it is expected that  ex-situ process provided  fairly good  junction in  Co/YBCO.
However, if it happens to have oxygen deficient surface of YBCO due to Co deposition, that may help in obtaining longer coherence length. Because, it has been reported that coherence length of YBCO increases with oxygen deficiency.\cite{ossa}
We have prepared samples with 50 x 70 $\mu$m$^{2}$  contact area by photo-lithography and selective area etching. The samples are made in a cross geometry. The measuring electrodes are fabricated by depositing silver contacts on the contact pads.
Two  contacts are made on Co pads, while the other two
contacts are made on YBCO pads.
The  resistance vs applied  bias voltage
measurement has been carried out  by AC modulation technique using a lock-in amplifier  in a continuous flow cryostat from 10 to 100~K. The applied frequency is 11 Hz. It has been observed that  the critical current density of YBCO has not
been changed after deposition of Co.\cite{ng}
 The junction resistance lies in the M$\Omega$ range at room temperature and it can be in tunneling limit.The reason of high resistance can be the rough interface of YBCO itself. Another possibility may be the formation of Cobalt oxide layer between YBCO and Co. Hence, we could not observe clear Andreev reflection. Nevertheless, we have observed  some abnormalities in AR and tried  to explain  them with the help  of modified BTK theory in the following sections.


\section{Results and Discussion}
We have observed resistance variation  at Co/YBCO   with 50 x 70 $\mu$m$^{2}$ contact areas at low temperatures within the superconducting gap ($\Delta$ = 20 meV) of YBCO. The measured normalized  resistance at  10, 15 and 100~K  are presented  in Figure~1.  It is noticed that a particular feature of the measured voltage with in the superconducting gap (Figure~1(a) and (b)) is not present at 100~K (Fig.~1(c)). Our study suggests that this feature  has been caused due to AR. However, even at 10-15~K the signature of AR is superimposed by  multi-peaks possibly due to the high junction resistance and large  rough interface area.
%
In this regard, it has to be mentioned that, we have not observed multi-peaks in AR for Co/YBCO samples of much smaller contact area. The  modified BTK modelling of the normalized resistance data of those samples reports a spin polarization of Co around 34$\%$.\cite{ng}\\

In order to explain the origin of multi-peaks in normalized resistance data of Co/YBCO with  50 x 70 $\mu$m$^{2}$ contact areas, we can proceed in the following way.
%
%
It is assumed  that a rough, large N/SC interface can be divided into several point contacts with different  barrier heights $Z$ and scattering rate $\Gamma$.
The barriers are distributed in space. Every barrier  corresponds to one point contact whose dimension  should be less than the mean free path of the FM and  coherence length of SC.
When the contact area is large, a shift of the zero bias resistance (ZBR)  dip in the differential resistance data can also occur.
In principle, the assumption of a delta function potential barrier across N/SC junction  in BTK model only applies for ideal junctions. For example, the roughness of real FM/high T$_{C}$  junctions can be taken into account by randomly distributed potential barriers at the interface.  These are uncorrelated random potentials which can result into local electric fields at the junction.
  The effect of electric field on ballistic transport of quasiparticles in mesoscopic system  has already been  analyzed theoretically. It is reported that a peak  can develop in AR conductance as a function of biased voltage and it can change sign when  the direction of electric field is changed.\cite{Mina}
A local electric field can alter the distribution of energy levels, which may change the density of states of quasi particles at the Fermi level. As a result, one can observe asymmetry in the normalized conductance vs voltage plot since this plot usually follows the density of states.
Hence, an additional voltage drop has been introduced in terms of  an extra energy loss in the expression of AR probability to account for the influence of local electric field.
This extra energy can be related to a shift of ZBR  dip in real experiments.
 Thus, we can assume  that when a current passes through a  single point contact at a real junction with rough, large interface, an additional voltage drop $\it{V_{ad}}$ depending on the magnitude of such  local field may  occur which can be  represented  by  an extra energy $\it{E_{ad}}$. The current  vs voltage characteristics have been numerically calculated with an additional energy loss  and are represented  in Figure~2(a).
 Different slopes are recognized inside and outside  the superconducting gap (20 meV). The slope change around 20 mV is due to
the  transition of  I vs U characteristics from  the AR  to normal Ohmic regime.
The  shift of the zero bias dip in the  differential resistance corresponds to the extra energy $\it{E_{ad}}$ (see Figure~2(b)). It should be pointed out that no change in the modelled  $\it{P}$ value occurs due to  $\it{V_{ad}}$.\\

 As described before in the case of a rough,large junction area interface,  several point contacts  may be formed at the FM/high T$_{C}$ junction and each contact may be influenced by a different local electric field causing extra potential drop.  Eventually, different contacts can give rise to  various resistance dips  due to Andreev reflection, shifted from the zero bias position, when the applied bias is  less than $\Delta$. This can be a possible cause of the so called noise as a function of bias voltage in  differential resistance measurements within the sub-gap region.
  Figure~3 describes the effect of multi-point contacts formed at a rough interface on R$_{N}$. If the total contact area is assumed to be formed by different separate point contacts with same length and area and different $\it{Z}$, $\it\Gamma$ and  $\it{V_{ad}}$ , the total normalized resistance will be the mean value (bold solid line in Figure~3(b) and (c)). Although the diameter of the total contact area is larger than the  mean free path and coherence length, the dimension of an individual point contact is still smaller than that scale. The randomness is realized by choosing the values of  $\it{Z}$, $\it\Gamma$ and  $\it{V_{ad}}$  arbitrarily.
 The number of  modelled point contacts increases from 1,2 to 5 in Figure~3(a), (b) and (c), respectively. The number of peaks in the sub-gap region increases with the number of separate point contacts.
 Note that the variation in mean normalized resistance amounts to 1-5$\%$ (Figure~3(c)) compared to 12$\%$ for an ideal contact (Figure~3(a)).
 This analysis explains  why there are multi-peaks  in  experimental normalized resistance data (Figure~1) taken on Co/YBCO  50x70 $\mu$m$^{2}$ samples.
 Moreover, the analysis also points out the reason of  much smaller variation in normalized resistance for  Co/YBCO  50x70 $\mu$m$^{2}$ samples (Figure~1)  than  that for 1x1 $\mu$m$^{2}$  samples.\cite{ng}
  Our  modelling  implies  that $\it\Delta$ and $\it{P}$  remain unchanged if several point contacts form the total junction area.
  The anisotropy of d-wave order parameter  is not considered in our modelling. Because, the contact direction is oriented along   c  crystallographic axis, where  anisotropy does not  play much important role as described by Barash $\it{et\
al.}$ \cite{Barash}.
 However,  since the contact  area is big , there may be some effect of   anisotropy.
 We have not considered the special dependence of energy gap ($\it\Delta$) near interface, because it is assumed that the  composition of YBCO surface is constant all over the junction area.  Hence, $\it\Delta$ value is same for every point contact.
  Nevertheless, proximity induced spatial dependence of gap parameter has been taken in to account for each contact, since two values of delta ( $\it\Delta_{1,2}$) are considered.\\

 We have shown  representative plots of  experimental and modelled normalized resistances for a Co/YBCO  50 x 70 $\mu$m$^{2}$ sample at  10 and 15~K  in Figure~4. It can be pointed out that our modelled normalized resistance can follow the original experimental data. The values of $\it{P}$ (30$\%$) and $\Delta_{1,2}$(27 meV) are also very close to those reported in the literature.\cite{tedro} We kept $\Delta_{1}=\Delta_{2}$ here, since proximity effect is not important for a FM/high-T$_{C}$ junction like Co/YBCO. Actually, ferromagnetic material (like Co) breaks the cooper pairs.
 If cooper pairs can not be formed in Co, proximity layer will not exist.
 But,  there is some difference between modelled and experimental  normalized resistance in magnitudes  along the vertical axis. This enhancement in normalized resistance and occurrence of extra peaks  can be caused by the critical current $\it{I_{C}}$ which will be discussed in the next section.\\

\subsection{Effect of critical current}
There are  additional resistance peaks in AR data at a bias larger than the superconducting gap (eV ${>}$ $\Delta$).  This is not predicted by BTK theory.\cite{goutam} Furthermore,  an  anomalous enhancement in the sub-gap conductance  (eV ${<}$ $\Delta$) in
nanoscale pinhole junctions  is observed. This may  sometimes become significantly larger than the conductance doubling expected from the point contact AR process.\cite{hind}
When the region of the FM/high T$_{C}$  junction is driven into the normal state due to current-induced pair breaking as the current becomes larger than $\it{I_{C}}$,  additional  peaks in differential resistance may be measured. If the contact current density depending on the  effective contact area increases above the critical current density $\it{j_{C}}$, the superconductor is transformed into a normal conductor.
It is possible to estimate the effective area of a  N/SC junction from its normal state of conductance $G_{N} \approx \frac {e^{2}}{ \hbar} k_{F}^{2}\mathcal{A}^{'}$, where k$_{F}$ is the Fermi vector of the superconductor and $\mathcal{A}^{'}$ is the effective contact area. For example,  k$_{F}$ is  1 x 10$^{8}$ cm$^{-1}$ for YBCO and in normal state  the contact resistance amounts to  1M$\Omega$  in case of $\mathcal{A}$= 50x70 $\mu m^{2}$ Co/YBCO or Zn(Al,Co)O/YBCO  junction, and  $\mathcal{A}^{'}$  amounts to 40 x 10$^{-6}$ nm$^{2}$.
Note that the effective contact area ($\mathcal{A}^{'}$) is much smaller than the total area ($\mathcal{A}$) in this kind of high resistance FM/high T$_{C}$ junctions. If for example a 40 pA current  passes across the junction, the current density ($\it{j}$) will be around 1x10$^{8}$ A/cm$^{2}$, whereas the $\it{j_{C}}$ for YBCO lies around 3x10$^{7}$ A/cm$^{2}$ at 4.2~K.\cite{beck} Hence, it is important to take into account the effect of critical current in the modelling. In order to do that, we have subdivided the total  junction current into I$_{Crit}$  and I$_{AR}$ in the full energy range. The calculated I vs U curves are shown in Figure~5                                                                                           (a). The slope in and outside the sub-gap region depends on the I$_{Crit}$ / I$_{AR}$ ratio.
 Similarly, the junction resistance is divided into two series resistances. One  is  R$_{AR}$ which develops as the AR current  passes across the junction. The other is R$_{Crit}$ which comes into play when the  junction current crosses the limit  $\it{I_{C}}$.  Simulation has been carried out considering the effect of the two subdivided junction resistances for the whole energy range and keeping the ratio of  R$_{Crit}$/R$_{AR}$ as a controlling parameter. It can be seen in Figure~5(b)  how the increase of  R$_{Crit}$/R$_{AR}$ creates enhancement in  resistance peaks in the sub-gap region.  For  R$_{Crit}$/R$_{AR}$=0 the plot agrees well with the prediction of BTK theory. It is clearly seen that with increasing  R$_{Crit}$/R$_{AR}$ ratio, the enhancement of   resistance (or decrease of conductance)  is more than the usual doubling described by  the BTK model for ideal Andreev reflection.\\
\section{Conclusion}
In conclusion, we have  modelled the normalized resistance data of realistic FM/SC  junctions. We have modified  the original BTK formula with incorporation of proximity effect $\it{\Delta_{1,2}}$ , spin polarization  $\it{P}$ and  broadening parameter for finite life time effects and interface scattering $\it{\Gamma}$. We have demonstrated the occurrence  of multi-peaks in  normalized resistance data for Co/YBCO sample with 50x70 $\mu$m$^{2}$ contact area. In order to explain the multi-peak nature of AR, an additional  voltage drop $\it{V_{ad}}$ in terms of extra energy loss $\it{E_{ad}}$ due to  local electric fields at  rough interfaces  is included in the modified model. This accounts for the shift of  the zero bias resistance dip in R/R$_{N}$. Furthermore, we have explained the anomalous enhancement of normalized resistance due to  the crossing of critical current limit across a  FM/SC  junction. The presented modified BTK formula may be used to model resistance peaks in the sub-gap region which have been experimentally observed at rough, large area FM/high T$_{C}$ junctions. Essentially, the model is useful because it can help to extract the value of spin polarization  $\it{P}$ for FM and  the superconducting gap parameter $\it{\Delta}$ for SC from  noisy differential resistance measurement data  caused by junction interface inhomogeneities.

\subsection*{Acknowledgement}
 Partial financial support  from Alexander von Humboldt foundation (N.G, INI 1120445 STP) and from BMBF (H.S, FKZ 03N8708)  is gratefully acknowledged.\\

 $^{*}$ Corresponding author:E-mail:  ghosh.nilotpal@gmail.com, present address:Superconductivity Research and Application Section, Material Science Division, Indira Gandhi  Centre for Atominc Research , Kalpakkam-603102, India

\subsection*{ Figure and Figure Captions}

%
%
%

\begin{figure*}[h]
\includegraphics[width=10cm]{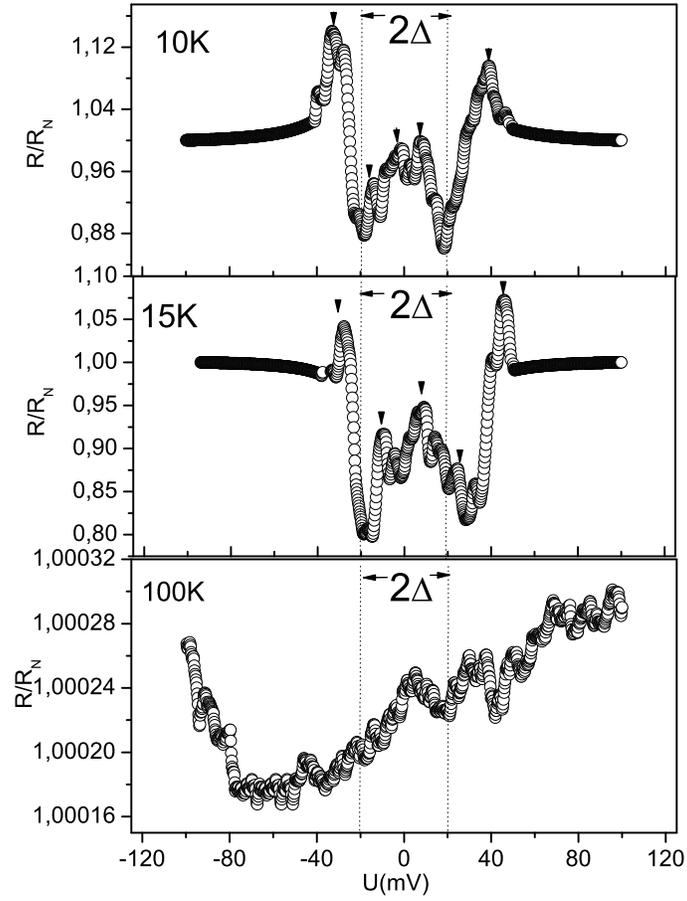}
\caption{ Points are   normalized resistance  for Co/YBCO  50 x 70 $\mu$m$^{2}$ sample at
 (a)  10~K, (b)  15~K and (d)  100~K. Two vertical dashed lines are to guide the region of superconducting gap 2$\Delta$ for YBCO. The arrows are to point out the additional peaks in normalized resistance data. It is clearly seen that some features of  AR are superimposed by multi-peaks in the
 sub-gap region, which is not observed at 100~K.}
\end{figure*}

\begin{figure*}[h]
\includegraphics[width=10cm]{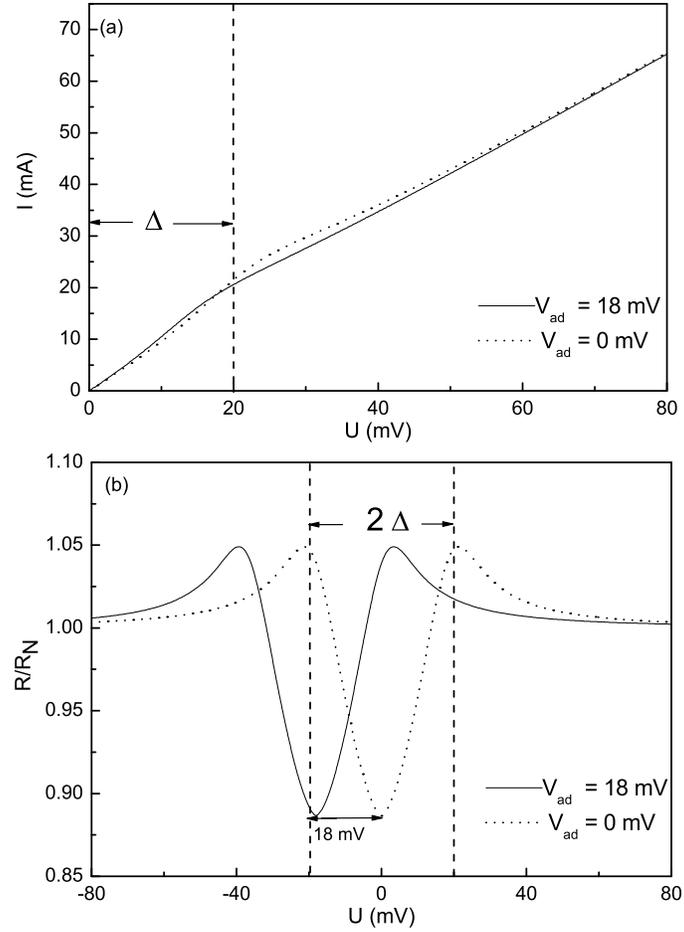}
\caption{(a) Modelled current vs voltage curves and (b) normalized resistance vs voltage curves at 10~K . The curves are calculated using  $\it{Z}$ = 0.45, $\it\Delta_{1}$= 17 meV, $\it\Delta_{2}$= 20 meV, $\it{P}$ = 30$\%$, $\it\Gamma$ = 7 meV and an additional voltage $\it{V_{ad}}$ given in the legend. The difference between the two  curves is clearly visible in the corresponding normalized resistance curves (b). Vertical dashed lines limit the superconducting gap for YBCO, which is 20 meV. }
\end{figure*}

\begin{figure*}[h]
\includegraphics[width=10cm]{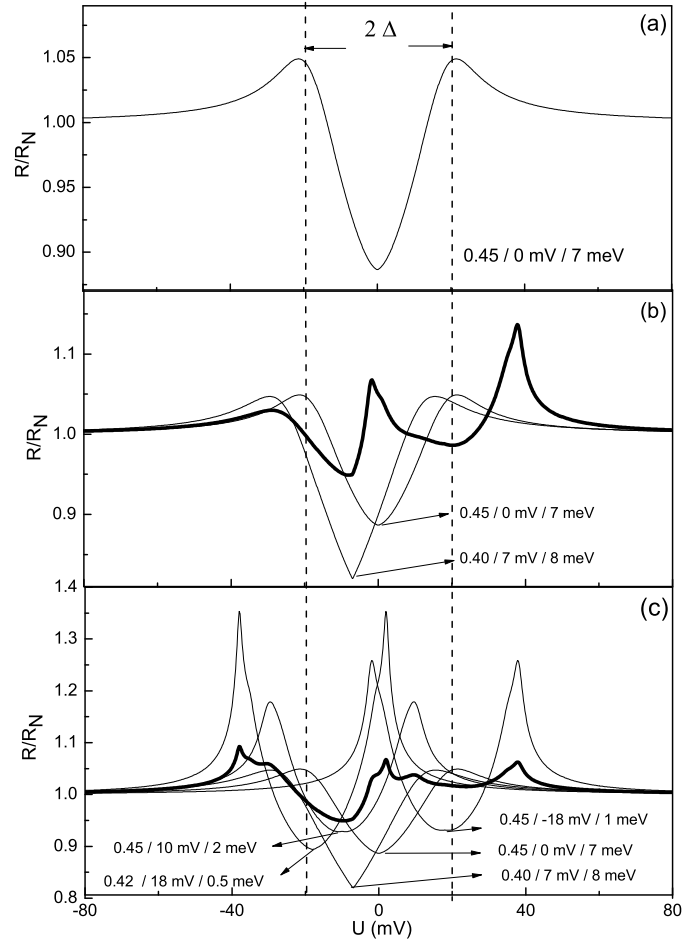}
\caption{ Modelled normalized resistance R/R$_{N}$ versus U curves at 10~K using the same $\it\Delta_{1}$ = 17 meV,$\it\Delta_{2}$= 20 meV and  $\it{P}$ = 30$\%$, but different $\it{Z}$, $\it\Gamma$ and  $\it{V_{ad}}$.  The $\it{Z}$/ $\it{V_{ad}}$ /$\it\Gamma$  values  are given in the legends and are (a)  0.45/ 0 mV / 7 meV (b)  0.45, 0.40/ 0,7 mV /7, 8meV and (c)  0.45, 0.42, 0.40/  10, 18, -18, 0, 7 mV/ 2, 0.5, 1, 7, 8 meV. It may be considered that there can be multi-point contacts at the rough interface of FM/high T$_{C}$. Each corresponds to one solid line in the plot. If the total contact area is assumed  to be formed by different  separate point contacts with similar area, its total normalized resistance is the mean value (bold solid line). Since each of these contacts is assumed to have a different potential barrier, local electric fields may cause various shifts in the zero bias resistance dips corresponding to different $\it{V_{ad}}$. As a result, there will be multi-peaks in the sub-gap region causing unavoidable noise. Vertical dashed lines limit the superconducting gap for YBCO, which amounts to  $\it\Delta$ = 20  meV.  }
\end{figure*}

\begin{figure*}[h]
\includegraphics[width=10cm]{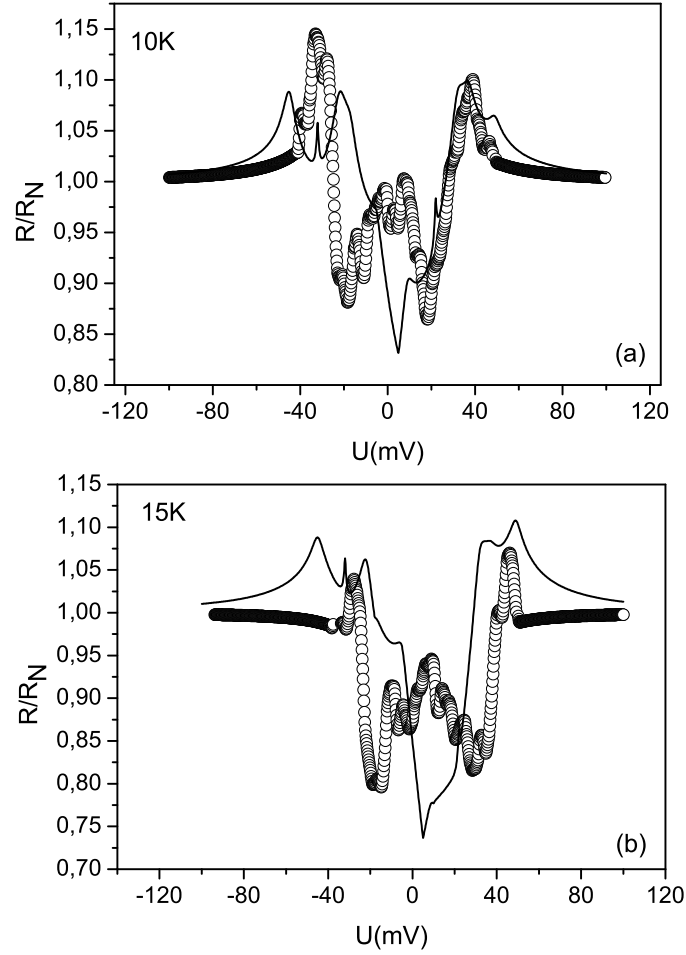}
\caption{Points are experimental normalized resistance  for a Co/YBCO  50 x 70 $\mu$m$^{2}$ sample at 10 and 15~K. The bold solid line is the modelled normalized resistance considering four different point contacts with  $\it{Z}$/ $\it{V_{ad}}$ /$\it\Gamma$  values as (a)0.075,0.45,0.5,0.5,0.93 / -5,-22,18,-10,5 mV/ 2.2,2,2.1,2.2,0.3 meV   (b) 0.075,0.065,0.06,0.05, 0.93 / -5,-22,18,-10,5 mV/ 2.2,2,2.1,2.2,0.3 meV  for 15~K respectively. The $\Delta_{1,2}$ and $\it{P}$ are kept as 27 meV and 30$\%$.It is noticed that the simulated plots  follow the original experimental data with multi-peaks. But there are still some differences in the magnitudes of experimental and modelled  normalized resistance along the vertical axis. Hence, we presume this extra resistance peaks originate from the effect of critical current.}
\end{figure*}

\begin{figure*}[h]
\includegraphics[width=10cm]{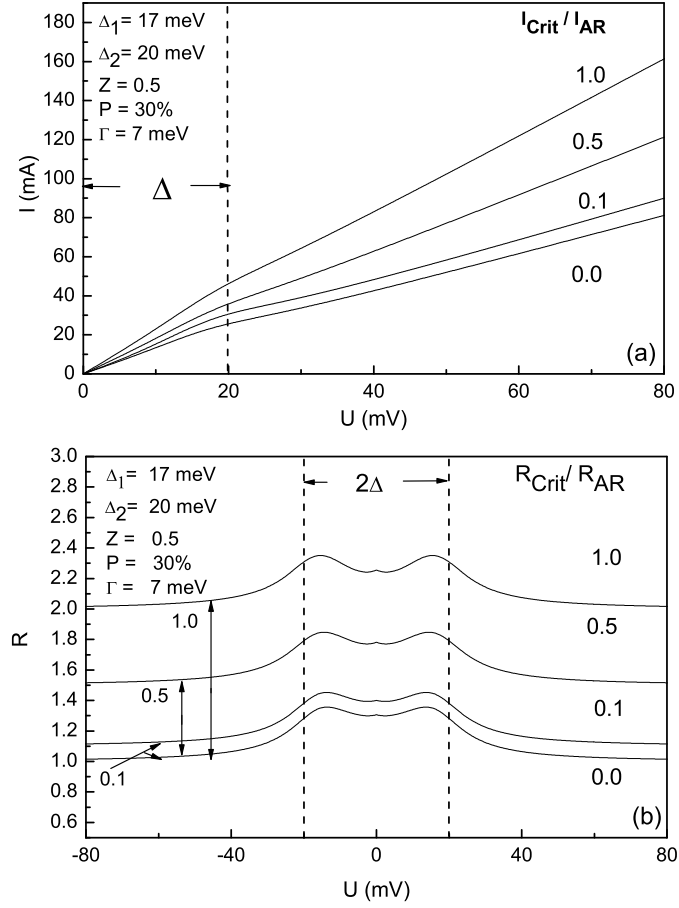}
\caption{ (a) Calculated   I vs U  curves with different I$_{Crit}$ / I$_{AR}$  ratios keeping the AR current constant and fixed $\it\Delta$, $\it{P}$ and $\it\Gamma$ as shown in the legends. The slope increases with the I$_{Crit}$ / I$_{AR}$ ratio (b) Modelled resistance R  with incorporated effect of critical current. The total junction resistance is composed of  R$_{AR}$ for AR current and R$_{Crit}$ for critical current in the full energy range. As the ratio of R$_{Crit}$/ R$_{AR}$ increases, the peak height at eV = $\Delta$ increases  more than expected by BTK theory. Note that the resistance curves ( R$_{Crit}$/ R$_{AR}$ $\neq$ 0) seem to have been shifted vertically from the position of the ideal BTK curve (R$_{Crit}$/ R$_{AR}$ = 0). The magnitude of  the shift for every curve  is similar to its R$_{Crit}$/ R$_{AR}$  ratio. Vertical dashed lines limit the superconducting gap for YBCO, which is 20 meV.}
\end{figure*}

\end{document}